\documentclass{emulateapj}
\usepackage{apjfonts}

\def\submitms{n}

\newenvironment{ffigure}[2][0]{
  \begin{figure}
    \def\longcap{#2\\}
}{
  \figcaption{\longcap}%
  \end{figure}
}

\newenvironment{ttable}[3][0]{%
  \begin{deluxetable}{#3}
    \tablecaption{#2}
    \tablewidth{0pt}
}{
    \enddata
  \end{deluxetable}
}
\let\oldtablehead\tablehead
\renewcommand\tablehead[1]{
  \oldtablehead{#1}
  \startdata
}

\newcommand*\thefigurefile[1]{#1}

\newcommand*\toup[1]{\uppercase{#1}}
\newcommand*\steps[1]{%
\if\submitms y
\subsection*{#1}
\else
\begin{center}
\textit{#1}
\end{center}
\fi
}

\shorttitle{Wavelet-Based Removal of Fringes}
\shortauthors{Rojo \& Harrington}

\slugcomment{Submitted to ApJ on November 14, 2005.  Accepted on May 20, 2006}

\begin{document}

\title{A Method To Remove Fringes From Images Using Wavelets}

\if\submitms y

  \author{Patricio M. Rojo}
  \affil{516 Space Sciences Bldg, Center for Radiophysics and Space
    Research \\ Cornell University, Ithaca, NY 14853-6801}
  \email{pato@astro.cornell.edu}
  \and
  \author{Joseph Harrington}
  \affil{326 Space Sciences Bldg, Center for Radiophysics and Space
    Research \\ Cornell University, Ithaca, NY 14853-6801}
  \email{jh@oobleck.astro.cornell.edu}

\else

  \author{
    P. Rojo\altaffilmark{1},
    and
    J. Harrington\altaffilmark{1},
  }
  \altaffiltext{1}{Center for Radiophysics and Space Research,
    Cornell University, Ithaca, NY 14853-6801 USA;
    pato@astro.cornell.edu, jh@oobleck.astro.cornell.edu.}

\fi

\begin{abstract}
We have developed a new method that uses wavelet analysis to remove
interference fringe patterns from images. This method is particularly
useful for flat fields in the common case where fringes vary between the
calibration and object data. We analyze the efficacy of this method by
creating fake flats with fictitious fringes and removing the fringes. We
find that the method removes 90\% of the fringe pattern if its amplitude
is equal to the random noise level and 60\% if the fringe amplitude is
$\approx 1/10$ of the noise level.  We also present examples using real
flat field frames. A routine written in the Interactive Data Language
(IDL) that implements this algorithm is available from the authors and
as an attachment to this paper.
\end{abstract}



\section{\toup{Introduction}}

The current class of telescopes with primary mirrors larger than $\sim$8
m in diameter allows researchers attainment of unprecedentedly high
signal-to-noise ratios. In addition, ever-increasing computer
capabilities have permitted quantitative analyses able to distinguish
trends weaker than the noise level. This has not only allowed
observations of fainter objects, but also observations of weak sources
spatially indistinguishable from a bright source, such as the spectrum
of an extrasolar planet orbiting a main-sequence star.  In such cases,
systematic errors that would not have been of importance when analyzing
the bright source are of concern when considering the fainter
source. Previously ignored systematic errors thus require algorithms
able to correct them.

One such systematic effect is the appearance of fringes in data arrays.
The strength of these fringes varies from instrument to instrument; we
have seen it range from less than a tenth of the noise amplitude to
roughly 5 times the noise amplitude.

Fringe correction methods found in the literature are either specific to
the instrument or assume a global fringe period (e.g.,
\citealt{met:malhiletal03, met:malhiletal03b, met:melwin95}).  The
latter correction type makes uses of Fourier filtering, a technique that
is less than satisfactory in the common case where the pattern's period
or amplitude varies over the image. Another common approach has simply
been to ignore the fringes in the hope that division of the debiased
data array by the flat field frame will eliminate the pattern, which was
supposed to remain the same.  However, flat fields and object frames do
not always share the same fringe pattern because flexure and variations
in the illumination geometry can change the pattern's amplitude or
period even on short timescales (Figs. \ref{flatchange} and
\ref{skychange}). Therefore, flat-field division could add a second set
of fringes rather than correct the first set.

\begin{ffigure}%
[Failure of flat field to correct fringes.]%
{Failure of flat field to correct fringes.
  Top: Debiased spectrum frame.  Bottom: Spectrum frame after flat field
  and bias correction. The fringe pattern is still visible. These
  images are subsections of a spectrum frame obtained with NIRSPEC at
  the Keck telescope. The horizontal white pattern is the spectrum's trace.
  Bright, white, vertical lines are the sky emission lines.
  \label{flatchange}}
\plotone{\thefigurefile{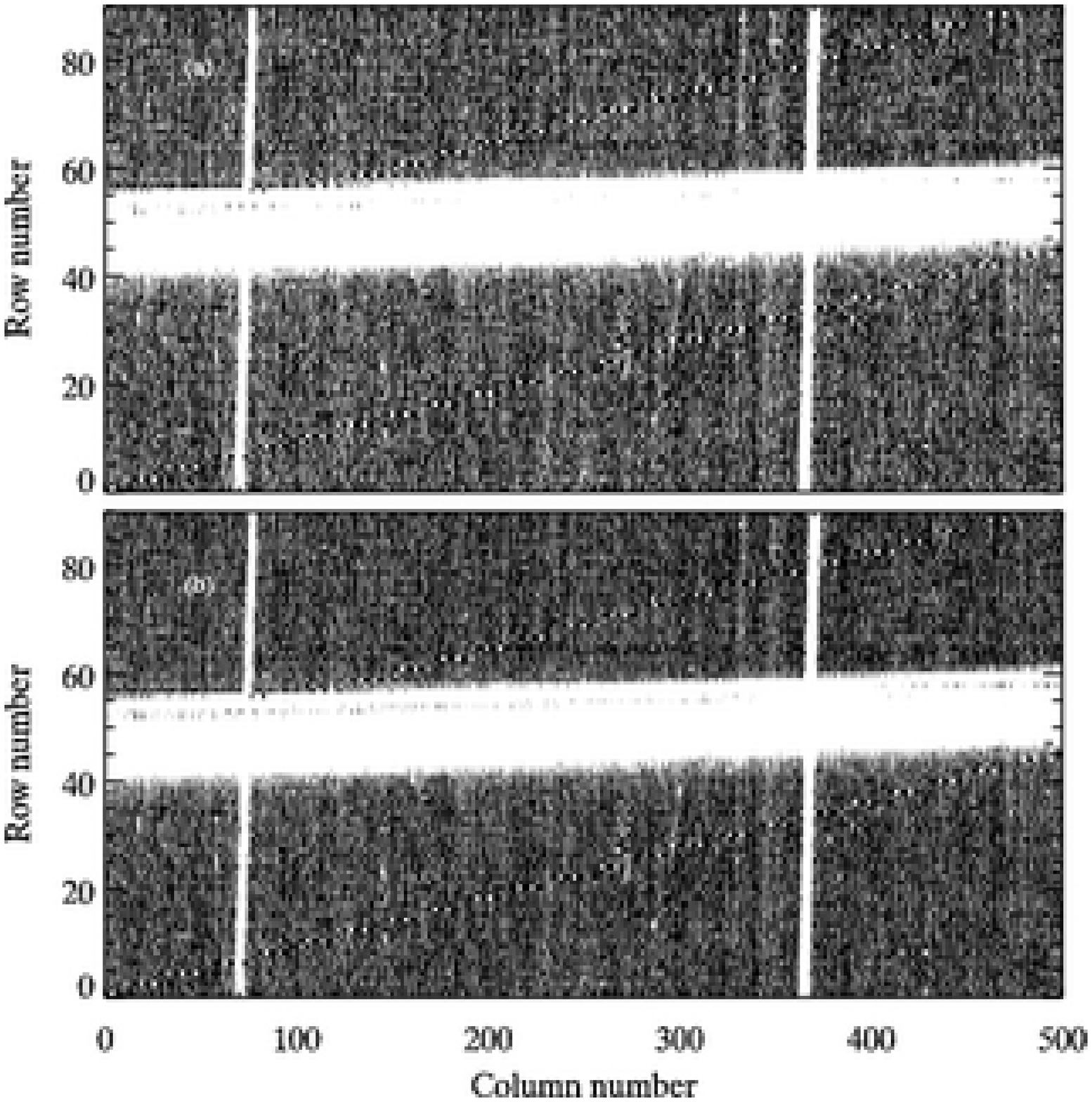}}
\end{ffigure}

\begin{ffigure}%
[Fringe variation in consecutive frames.]%
{Fringe variation in consecutive frames.
  Subsections of sky frames taken with NIRSPEC on the Keck II telescope
  (45 seconds integration). Fringe pattern can be seen varying between
  these frames (taken consecutively in order a, b, and c).
    \label{skychange}}
\plotone{\thefigurefile{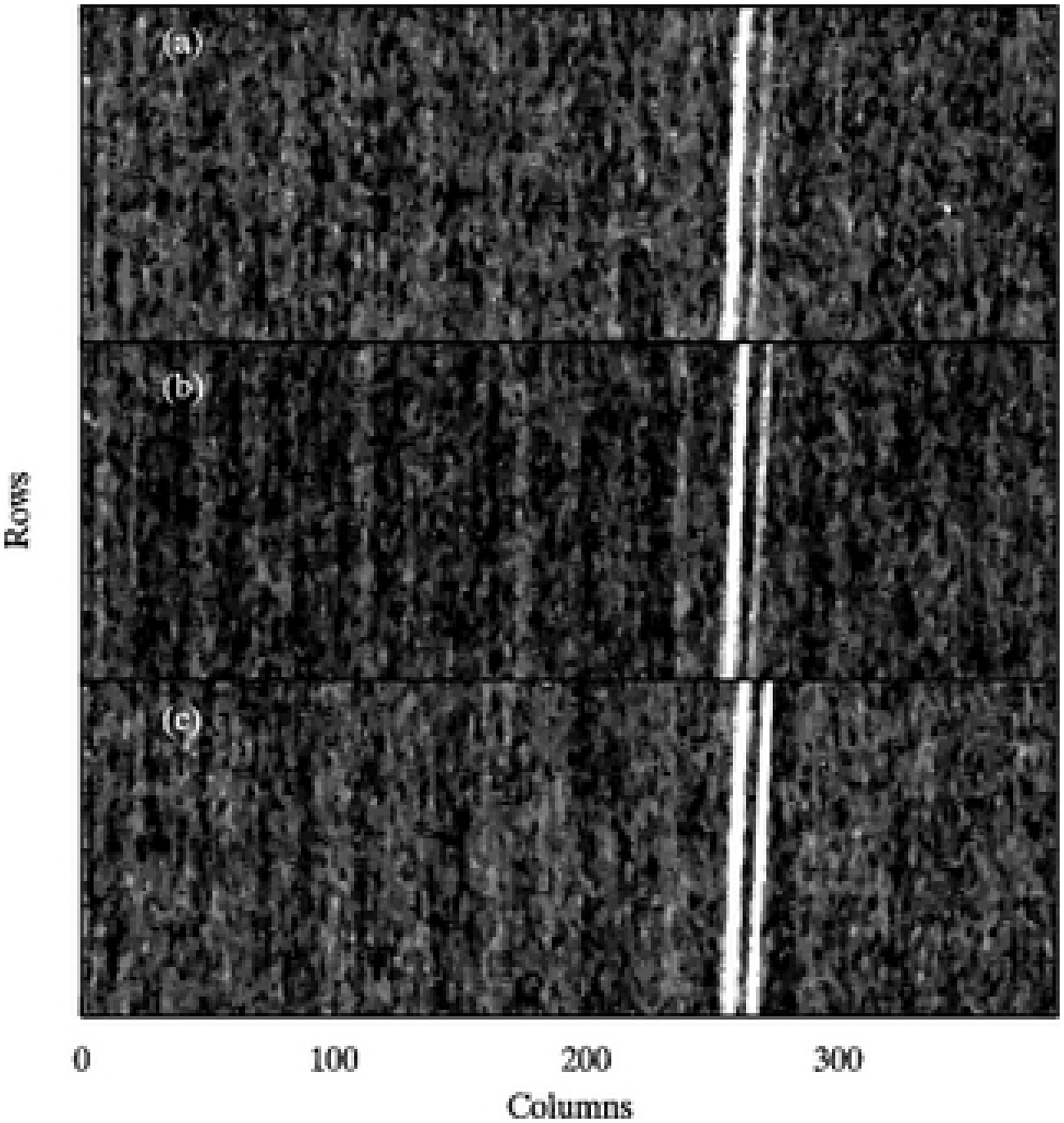}}
\end{ffigure}

Here we present an algorithm to clean two-dimensional (2D) arrays that
uses the wavelet transform, a local spectral technique (e.g.,
\citealt{b:stamur02,met:torcom98}\footnote{http://atoc.colorado.edu/research/wavelets/}).
Making use of the wavelet transform's linearity property, the algorithm
isolates the fringe pattern in wavelet space, does an inverse transform,
and then obtains a clean image by subtracting the reconstructed fringe
pattern.  The challenge is to do this correctly in the presence of
noise.  The algorithm presented here is not tuned to any specific
dataset and has been tested on flat field frames from ISAAC
\citep{instrument:isaac} at the Very Large Telescope (VLT) and NIRSPEC
\citep{instrument:nirspec} at the Keck telescope.

The algorithm also provides the framework for an extension to remove
fringes from object frames by interpolating the fringe pattern over
spectra or point sources.  Such an extension requires the design of
another algorithm to interpolate fringe parameters, which is beyond the
scope of the present work.  Thus, our method improves the quality of
extracted data when the fringes in the flat fields differ from the
fringes in the data, but will not solve the problem completely if the
fringes in the data array are significant. We show that there are many
cases where the fringes change between arrays obtained at different
times.

We implemented our algorithm in the Interactive Data Language (IDL, a
product of Research Systems, Inc., Boulder, Colorado).  The package (named
``defringeflat'') includes tutorial documentation. It is available under
the GNU General Public License from our
websites\footnote{http://www.das.uchile.cl/$\sim$pato/sw/ or
http://oobleck.astro.cornell.edu/jh/ast/software.html} or as an
electronic attachment to this paper.

Section \ref{fringe} gives a mathematical model of fringe
formation. Section \ref{algorithm} describes the algorithm.  Section
\ref{noise} discusses performance in the presence of noise.  Finally,
Section \ref{concl} discusses an applied example, summarizes the
benefits and limitations, and presents our conclusions.


\section{\toup{Fringes}}\label{fringe}

Fringes are produced by the interference of light reflecting between
parallel surfaces in an instrument.  They appear in many detectors of
visible and infrared light.  If we ignore multiple reflections, a
mathematical formulation \citep{b:rieke03} for the total intensity of
light ($I_a$) received on the position $x,y$ of the detector array is
given by
\begin{equation}
I_a(x,y) = I_n(x,y) + I_r(x,y) + 2\sqrt{I_n(x,y)I_r(x,y)}\cos\psi(x,y),
\end{equation}
where $I_n$ is the non-reflected intensity, $I_r$ is the reflected
intensity, and $\psi$ is the phase difference between the two beams. We
choose the $x$ coordinate such that
\begin{equation}
\psi(x,y) = 2\pi x/P(x,y)+\xi(x,y),
\end{equation}
where $P$ and $\xi$ are the period and phase of the fringe's pattern,
respectively.  Let $I_i$ be the incoming intensity before interaction
with the instrument. Then, $I_r(x,y)$ is proportional to the intensity
incident at a nearby position:
\begin{equation}
I_r(x,y)=a(x+\delta x,y+\delta y) I_i(x+\delta x,y+\delta y),
\end{equation}
where $\delta x$ and $\delta y$ are small displacements and the factor
$a$ includes reflectivity. Note also that $a\ll 1$.  If we can assume
that the incoming intensity field and the reflection geometry are
homogeneous on very short spatial scales, then $a(x+\delta x,y+\delta y)
I_i(x+\delta x,y+\delta y) \approx a(x,y) I_i(x,y)$.  On the other hand,
the non-reflected intensity ($I_n$) is proportional to the incoming
intensity ($I_i$) in the same coordinate, thus
$I_n(x,y)=b(x,y)I_i(x,y)$. Due to energy conservation, the
proportionality constant $b(x,y)$ is restricted by the previous
assumption to comply with
\begin{equation}
a(x,y)+b(x,y)=s(x,y),
\end{equation}
where $s(x,y) \lesssim 1$ accounts for losses due to scattering and
absorption.  Omitting the dependence on $x$ and $y$ for clarity, we
find for each position
\begin{eqnarray}
I_a &=& bI_i + aI_i + 2 I_i \sqrt{ab}\cos\psi \\
&=& I_i \left[s+2\sqrt{ab}\cos\left(2\pi x / P+\xi\right)\right]\\
&=& I_i \left[s+F\right],\label{recveq}
\end{eqnarray}
where
\begin{equation}
F=2\sqrt{ab}\cos\left(2\pi x / P+\xi\right)\label{frineq}
\end{equation}
is the oscillating fringe term. When interacting with the detector
array, Eq. \ref{recveq} is modulated to obtain the detected intensity
$I_d$.  Including detection noise, the modulation is given by
\begin{equation}
I_d = I_i \eta \left[s+F\right] + \epsilon,
\end{equation}
where $\epsilon$, which varies rapidly between pixels, includes all
noise sources and $\eta$ includes quantum efficiency, and pixel
collecting area, among other factors.  On the other hand, $\eta$ can be
decomposed as
\begin{equation}
\eta = \eta_0 + \eta_r,
\end{equation}
where $\eta_0$ is the smoothly varying component and $\eta_r$ is the
rapidly varying component, which includes uncorrelated differences
between the sensitivities of neighboring pixels. Typically, $\eta_r \ll
\eta_0$.  Bringing it all together, we obtain
\begin{equation}
  I_d = I_i \eta  s + \epsilon + I_i \eta_r F + I_i \eta_0 F.
\end{equation}

Our algorithm makes use of the linearity property of wavelets to find
and subtract the term $I_i \eta_0 F$, which is the predominant
contibutor at the period of the fringe pattern. The other terms will
only contribute in that frequency to a background level in the amplitude
of the wavelet transform. This background is considered in our algorithm
(see Step II in Section \ref{algorithm}).

Then, $\eta s$ can be corrected through flat-fielding to get the sought
intensity $I_i$ with a modified noise $\bar{\epsilon}$ given by
\begin{equation}
  \bar{\epsilon} = \epsilon + I_i \eta_r F.
\end{equation}
With typical values, $I_i \eta_r F \ll \epsilon$.


\section{\toup{Algorithm}}\label{algorithm}

The main steps in our procedure are listed in Table \ref{steps}.
Figs. \ref{ex_orig} -- \ref{ex_clean} illustrate the steps of the
algorithm using an example flat field.  Their captions contain details
regarding the example array, while the main text only refers to the
algorithm in general.  The example flat field is included in the
defringeflat package.

\begin{ttable}%
{Steps of Defringing Algorithm\label{steps}}%
{clc}
\tablehead{\colhead{Step} & \colhead{Description} &
  \colhead{Figure}}
 & Original image with fringe & \ref{ex_orig} \\
I & FOR EACH ROW & \\
 & $\Rightarrow$ Compute enhanced row & \ref{ex_enhrow} \\
& $\Rightarrow$ Compute wavelet transform &  \ref{ex_wave} \\
II & $\Rightarrow$ FOR EACH PIXEL IN ROW & \\
 & $\Longrightarrow$ Fit fringe transform's profile &
\ref{ex_profile} \\ 
III & FOR THE WHOLE ARRAY & \\
& $\Rightarrow$ Smooth fit parameters (optional) &
\ref{ex_smooth} \\ 
IV & FOR EACH ROW & \\
& $\Rightarrow$ Reconstruct wavelet array & \ref{ex_wave} \\
& $\Rightarrow$ Inverse transform & \ref{ex_wave},
\ref{ex_fringe} \\
& FOR THE WHOLE ARRAY & \\
& $\Rightarrow$ Subtract fringe pattern to obtain clean image &
\ref{ex_clean} \\
\end{ttable}

\begin{ffigure}%
  [Sample image with fringes.]%
  {Top: Sample image with fringes.  This
    flat field was obtained with the ISAAC instrument at the VLT.  Each of
    the numeric parameters indicated in the captions from
    Figs. \ref{ex_orig} to \ref{ex_clean} were found to be the most
    appropriate for this particular example, but will need to change for
    different images.  Columns 901--1024 and rows 0--149 and 951--1024 are
    vignetted and thus are cropped before analysis.  Periodicity can be
    estimated by eye at $\sim 40$ pixels in the center of the image.
    Bottom: Middle (512th) row.
    \label{ex_orig}}
  \plotone{\thefigurefile{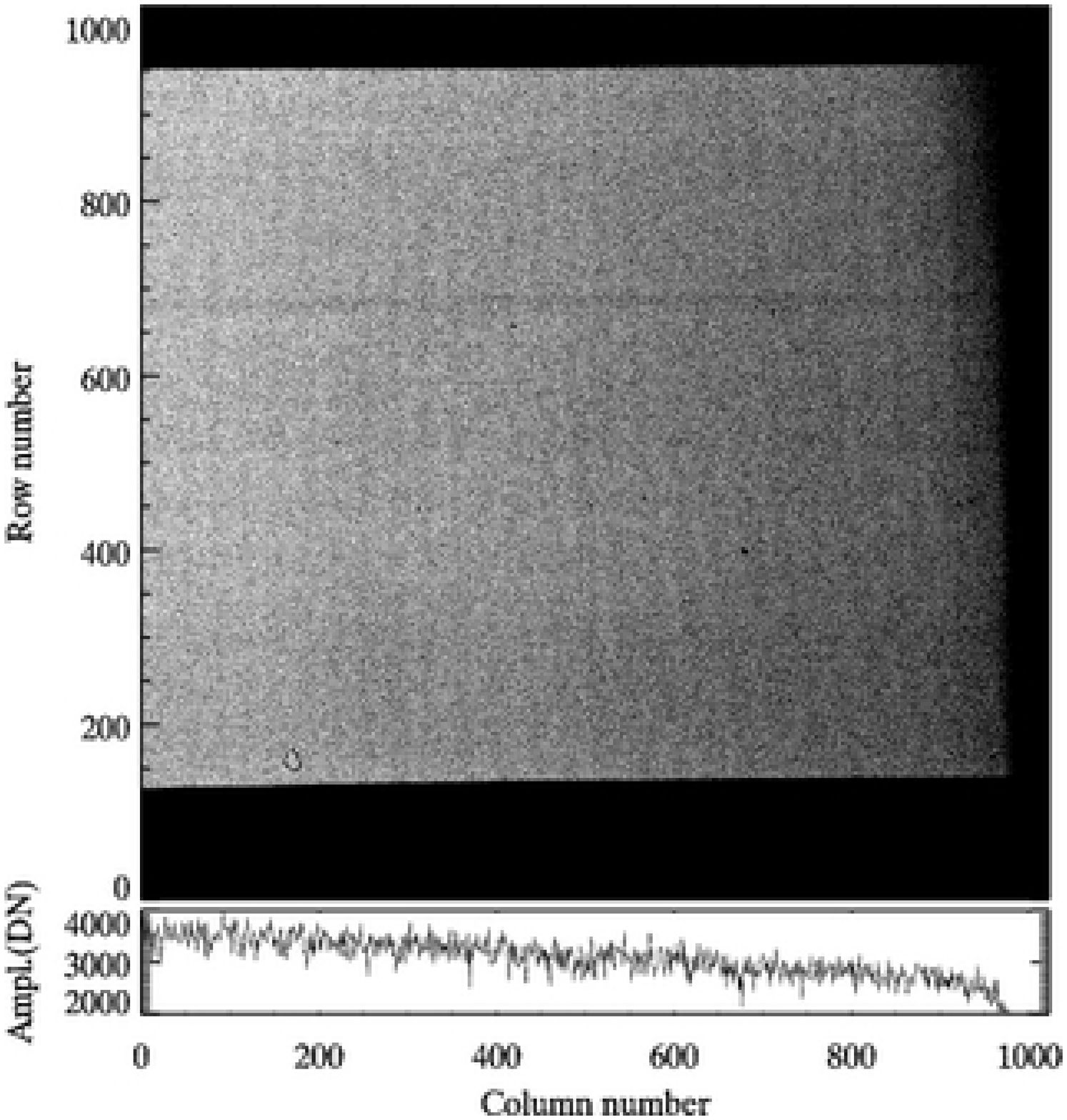}}
\end{ffigure}

All array borders whose values are not consistent with the image must be
cropped.  The fringes are allowed to have several different patterns,
which do not need to look like straight lines.  There are only two
requirements.  First, the period ($P$) of the fringe term
(Eq. \ref{frineq}) should change smoothly accross the array; and second,
only on a per-row basis, the period must be at least several pixels, but
it must also have at least a few oscillations per row.  To attain the
second condition it is acceptable to rotate the image by 90 degrees.
There are no constraints on how the phase $\xi$ can change accross rows
or the range over which $P$ can vary.  Hence, the algorithm can handle
many patterns that do not look like plane waves, such as patterns
resembling wood grain.

\steps{Step I. Enhanced Row and Wavelet Transform\label{ERwave}}
   For each image row we combine several surrounding rows to suppress
random noise and remove bad pixels. To do this, we replace each pixel in
the row with the median of a $1\times n$ sub-image centered on the pixel
and traversing $n$ rows (bin width, hereafter). We then subtract a
polynomial fit from the median-averaged row to obtain an {\em enhanced
row} (Fig. \ref{ex_enhrow}).  This subtraction significantly diminishes
the large-period (low-frequency) oscillations of each row (and their
corresponding wavelet amplitudes), allowing the next step to proceed
more efficiently.

\begin{ffigure}%
  [Enhanced rows]%
  {Top: Enhanced rows.  Each pixel of the
    array in Fig. \ref{ex_orig} is first replaced by the median average of
    the 41 closest pixels in the vertical direction.  A polynomial fit to
    each row is then subtracted.  The fringe pattern is enhanced and some
    bad pixels have been removed.  Note that the usable data area is
    reduced by 20 rows on the top and bottom because of the averaging.
    Bottom: Middle (512th) row.
    \label{ex_enhrow}}
  \plotone{\thefigurefile{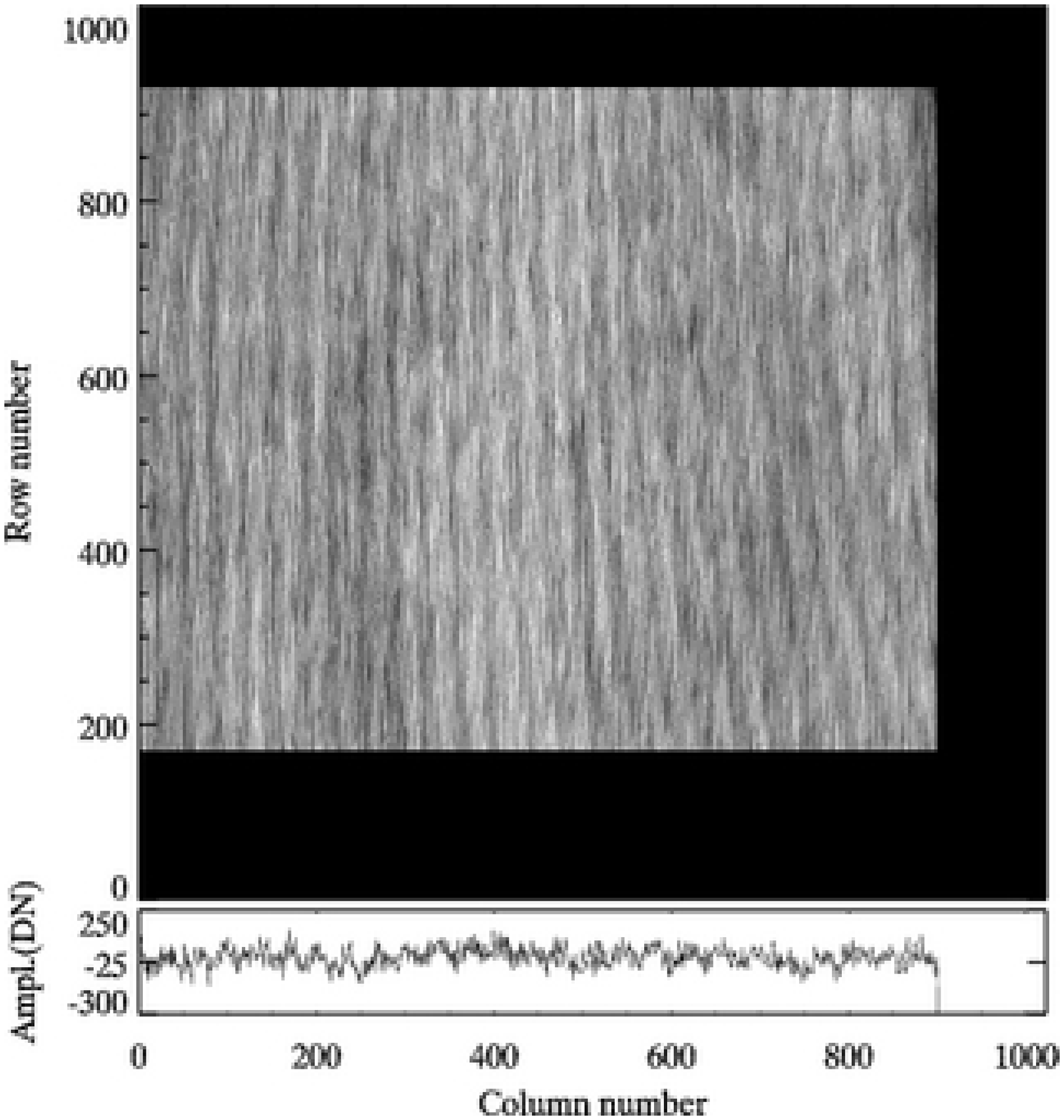}}
\end{ffigure}

We then compute the wavelet transform of each enhanced row.  The result
for each row is a two-dimensional, complex array, whose two dimensions
are column number and period.  There are several real and complex
wavelet bases to choose from, but Step II of this algorithm requires a
complex basis because real bases are not able to separate phase from
amplitude information.  For this particular example, we used the Morlet
wavelet because its functional form is the familiar quantum-mechanical
wave packet
\begin{equation}
\Psi(t) = \pi^{-1/4}e^{-t^2/2 - i \omega_0 t},
\end{equation}
which makes it well suited for smoothly varying periods. Here, $t$ and
$\omega_0$ are non-dimensional. For the Morlet basis, $\omega_0$ is the
only parameter; it dictates the minimum number of oscillations per row.
The Morlet basis also has the advantage of being compact in the
frequency domain.

However, the accompanying code allows the user to choose from several
other popular wavelets as they could be better suited for particular
data.  Steps II and III are computed over the complex array amplitudes
(wavelet array, hereafter).  The phases of the complex array must be
stored for use in Step IV.

\steps{Step II. Parametric Fit of Fringe Transform\label{trace}}

At the period of the fringe pattern, the wavelet array will contain a
prominent {\it fringe transform} pattern traversing the columns.  Its
amplitude depends on the amplitude of the fringe pattern (Fig.
\ref{ex_wave}).  This algorithm's success rests on our ability to
distinguish this feature from the background noise level of the wavelet
array.  The fringe transform may vanish for particular columns, but it
should be clearly distinguishable in most of each wavelet's array.
Improved sampling in period can be obtained by interpolation or by
decreasing the spacing between discrete scales in the wavelet
transform. A compromise should be chosen; the latter approach is more
accurate but demands more computer resources.

\begin{ffigure}%
  [Wavelet transform of an enhanced row.]%
  {Wavelet transform of an enhanced row.  In
    the center plots, the dotted line marks the cone of influence (COI);
    wavelet values below this boundary should not be trusted.  The dashed
    line shows the fitted trace.  Plot a: Middle enhanced row of
    Fig. \ref{ex_enhrow}.  Plot b: Amplitude of the Morlet wavelet
    transform of plot (a).  The wavelet is interpolated in period by a
    spline from the period sampling of the transform, and the fringe
    transform, a coherent pattern corresponding to a fringe with a period
    of 35 pixels, is clearly visible.  Plot c: Reconstructed fringe
    transform using a Gaussian fit (cf. Fig. \ref{ex_profile}).  Plot d:
    Fringe pattern after applying an inverse wavelet transform to plot
    (c), plotted over the input data.\label{ex_wave}}
  \plotone{\thefigurefile{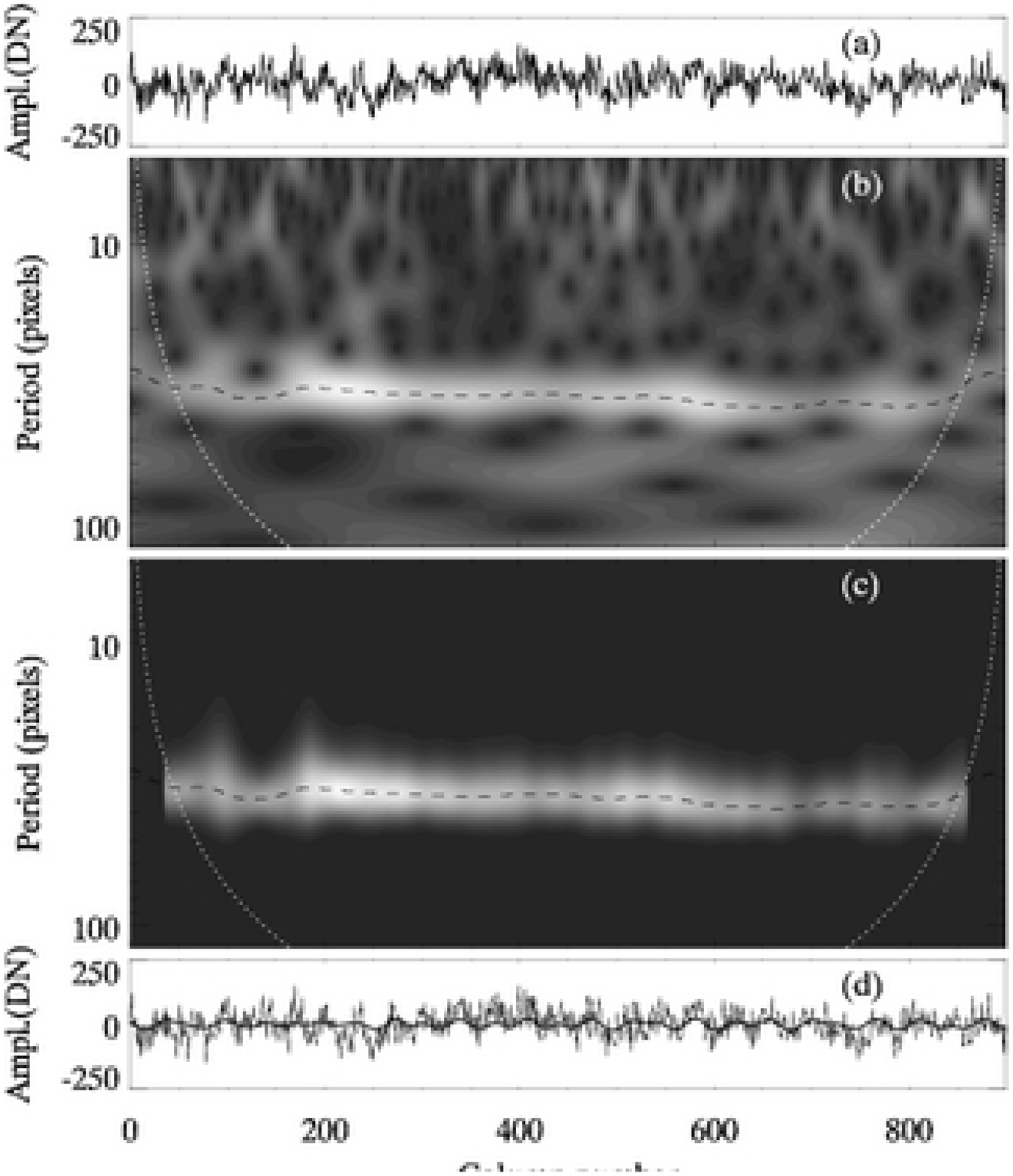}}
\end{ffigure}

We next either extract or fit the fringe transform's amplitude {\em vs.}
period for a given column (Fig.  \ref{ex_profile}).  Starting from a
reference column, the fringe profile is isolated by finding the first
local minima on both sides of the reference period.  Then one method is
chosen to represent the profile within the minima: either we use the
actual data within the minima ({\it trueshape} method, hereafter) or a
parametric function can be fit to the profile.  Only the latter approach
will allow execution of the optional Step III.  The value of the profile
must be zero outside the fringe transform.  Inside, on the other hand,
it is recommended that the fringe transform profile exclude a background
level (attributable to non-fringe image components, see discussion in
Section \ref{fringe}).  The highest point in the profile is used as the
new reference period for the next column.  The procedure is repeated for
the whole fringe transform, extending in both directions from the
reference column to the cone of influence (COI) boundary, beyond which
the wavelet values are significantly contaminated by edge effects.

\begin{ffigure}%
  [Cross--section along a column of the wavelet array]%
  {Cross--section along a column of the
    wavelet array. Top: Components diagram. Crosses are the amplitude of
    the wavelet points. The data points surrounding the region labeled
    as ``fringe transform'' are exactly what the trueshape method would
    reconstruct or what the parametric functions will fit. The region
    labeled ``fringe transform background'' is attributable to
    non-fringe components; in some of the tests it was used for the
    wavelet reconstruction (keep) and in some it was not
    (nokeep). Bottom: Gaussian fit (with background) to the region
    between the minima.  Crosses are as above, solid line is the fitted
    profile.\label{ex_profile}}
  \plotone{\thefigurefile{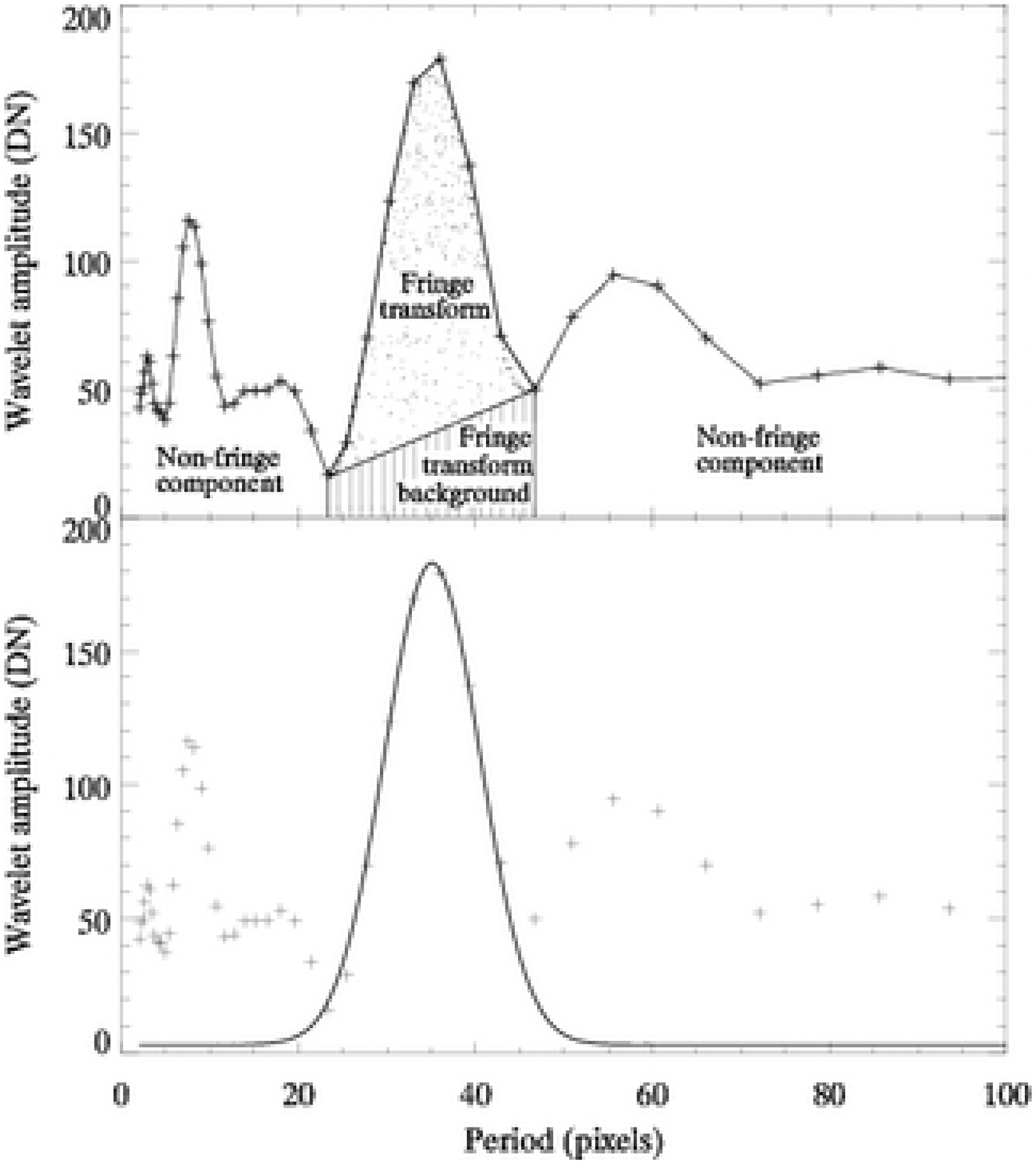}}
\end{ffigure}

To fit the profile we have experimented with plain Gaussian fits with
variable center (GVC) and Gaussian functions in which the center is
fixed at the maximum height (GFC).  Both Gaussian alternatives were
considered without a constant background parameter ({\it noback}), and
with this parameter.  In the latter case, the background value can be
kept or not when reconstructing ({\it keep} and {\it nokeep},
respectively).  In total, we have implemented six parametric fitting
methods (that can be smoothed or not in step III) and two trueshape fits
(nokeep and keep), for a total of 14 fitting methods.  The Gaussian
shape was chosen not only because it is a natural choice to fit a peak,
but also because it is the frequency-domain representation of the Morlet
wavelet.  The relative fringe-removal efficiency of these fits and of
trueshape is discussed in Section \ref{noise}.

\steps{Step III. Optional Parameter Smoothing\label{smooth}}

If a functional parametric fit was used in the previous step, one can
reduce the effects of noise by forcing the reconstructed fringe's
parameters to vary smoothly.  After repeating Steps I and II for every
row, a 2D array is obtained for each of the fitted parameters.  First,
we ``patch'' each of the parameter arrays by finding outliers beyond a
given number of standard deviations from the neighborhood median and
replacing them by that median value.  Then, we smooth the array with a
boxcar filter.  Figure \ref{ex_smooth} shows an example.

\begin{ffigure}%
  [Gaussian height parameter smoothing]%
  {Gaussian height parameter smoothing.  Top: Gaussian fit
    height parameter for central portion of example image.  Middle:
    Parameter after replacing all values more than $\pm1.5\sigma$ from
    the local median level with that level (patched array).  Bottom:
    Patched array after smoothing with a 19-pixel boxcar filter.  This
    procedure is repeated for each of the other Gaussian fit parameters.
    \label{ex_smooth}}
  \includegraphics[height=0.7\textheight]{\thefigurefile{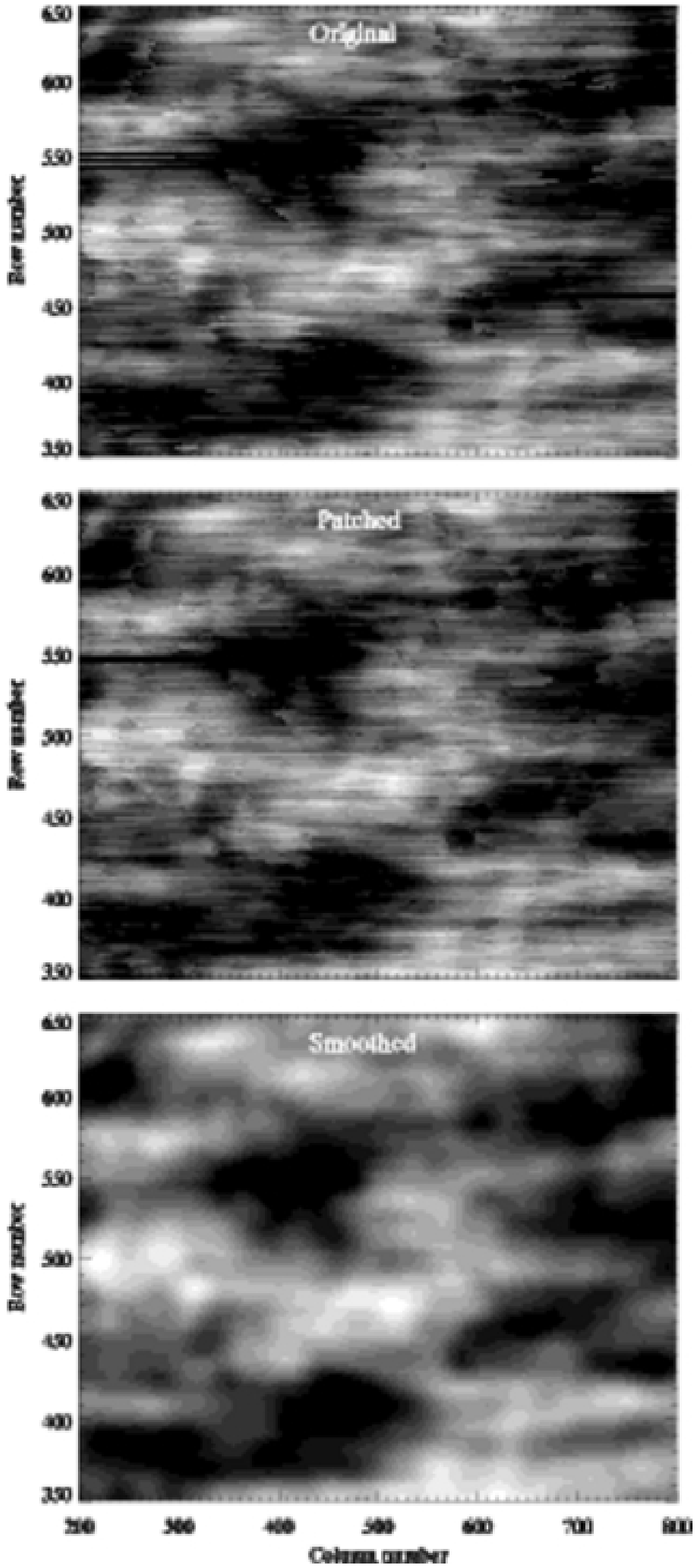}}
\end{ffigure}

\steps{Step IV. Reconstruction of Fringe Pattern\label{recon}}

We next evaluate the parameters to obtain the fringe's wavelet
amplitudes (Fig. \ref{ex_wave}).  Far from the reconstructed fringe
transform the amplitude must be zero because any non-zero value there
will cause unwanted noise in the reconstructed fringe.  In particular,
if a keep method is chosen, the reconstructed amplitude is set to zero
outside the local minima.  Finally, we apply an inverse wavelet
transform to the reconstructed wavelet amplitude and the corresponding
complex phases (see Step I.)

We repeat these steps for every row to obtain the image's isolated
fringe pattern (Fig. \ref{ex_fringe}).  Due to the optional smoothing,
the method to obtain the enhanced rows, and the COI boundary, the
recovered fringe pattern will have smaller borders than the original
image.  The fringe pattern can now be subtracted from the original image
(Fig. \ref{ex_clean}). Figure \ref{keckdefringe} shows another example
of this algorithm for a flat field from NIRSPEC at Keck.

\begin{ffigure}%
  [Reconstructed fringe pattern.]
  {Top: Reconstructed fringe pattern
    (cf. bottom panel of Fig. \ref{ex_wave}).  Bottom: Middle (512th)
    row.
    \label{ex_fringe}}
  \plotone{\thefigurefile{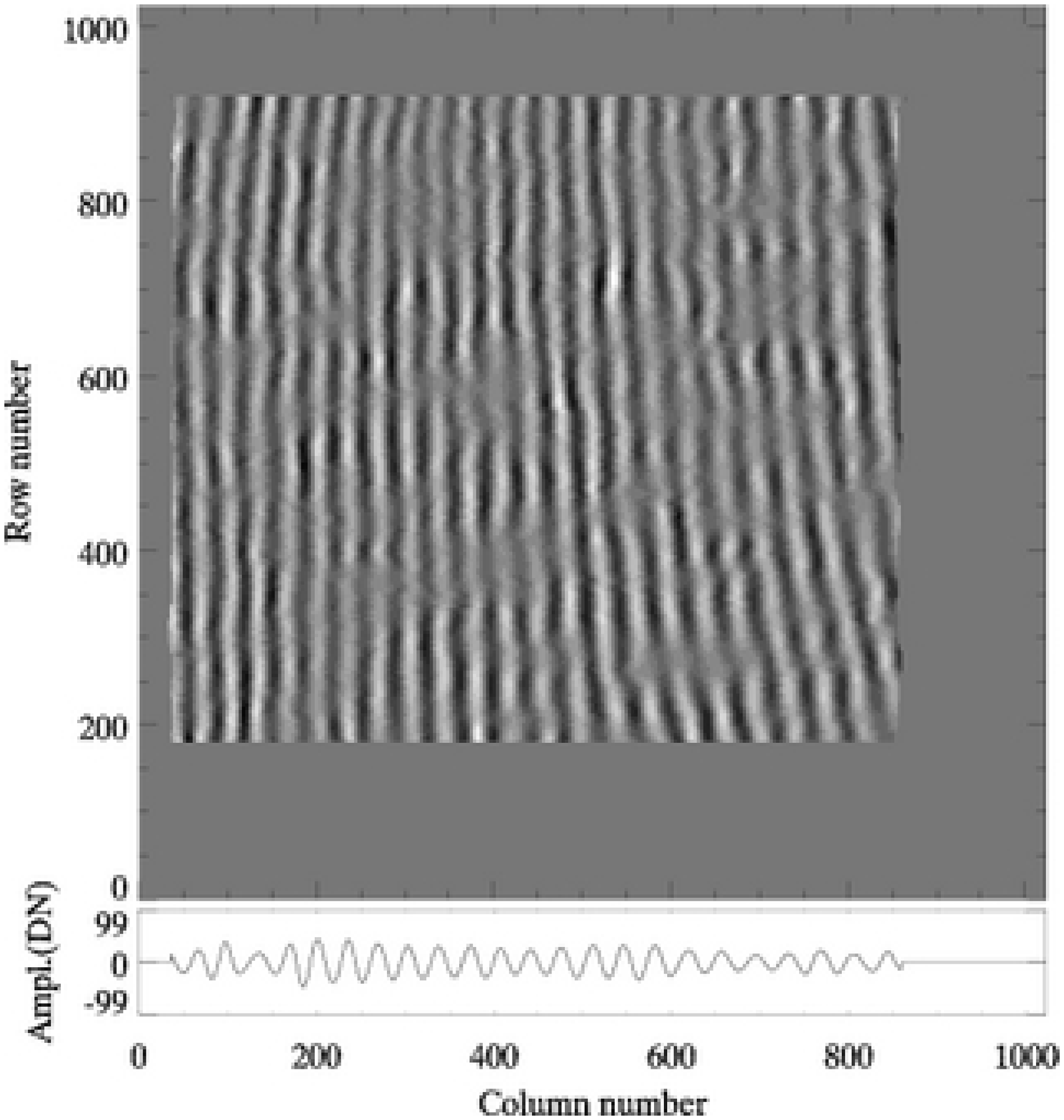}}
\end{ffigure}

\begin{ffigure}%
  [Cleaned image.]%
  {Top: Cleaned image. Flat field of
    Fig. \ref{ex_orig} minus the fringe pattern of Fig. \ref{ex_fringe}.
    Note that some of the edges remain uncorrected (see text).  Bottom:
    Middle (512th) row.
    \label{ex_clean}}
  \plotone{\thefigurefile{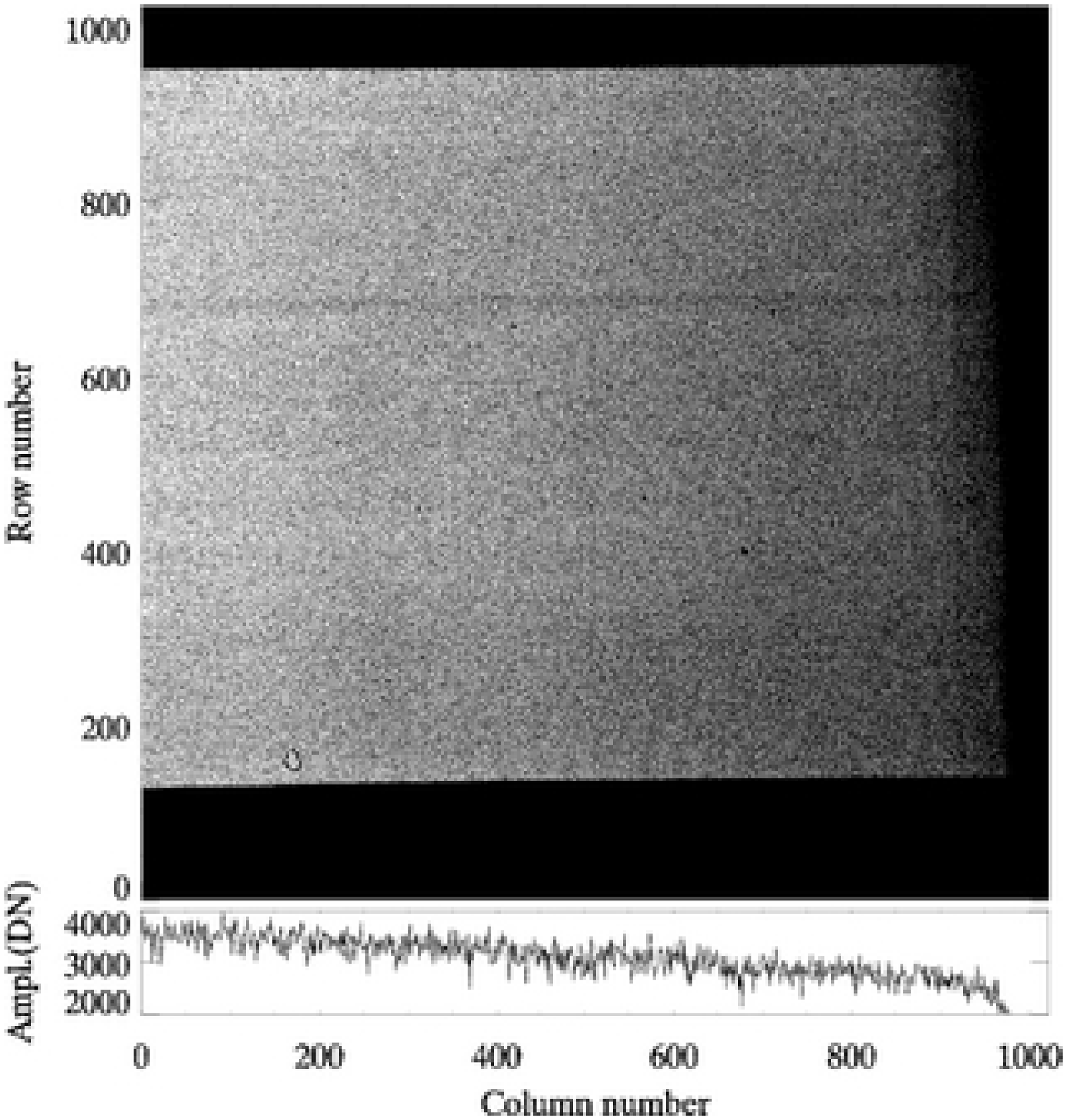}}
\end{ffigure}

\begin{ffigure}%
  [Example of fringe removal from a second instrument.]%
  {Example of fringe removal from a second instrument. Frames present a
    portion of a debiased flat-field frame from one order of the
    high-resolution NIRSPEC spectrograph at the Keck II telescope.  Top:
    Original flat field.  Bottom: Same flat field defringed by our
    method.  The algorithm had to be applied twice; once for a fringe of
    period $\sim20$ pixels and then for a fringe of period $\sim40$
    pixels.  \label{keckdefringe}} \plotone{\thefigurefile{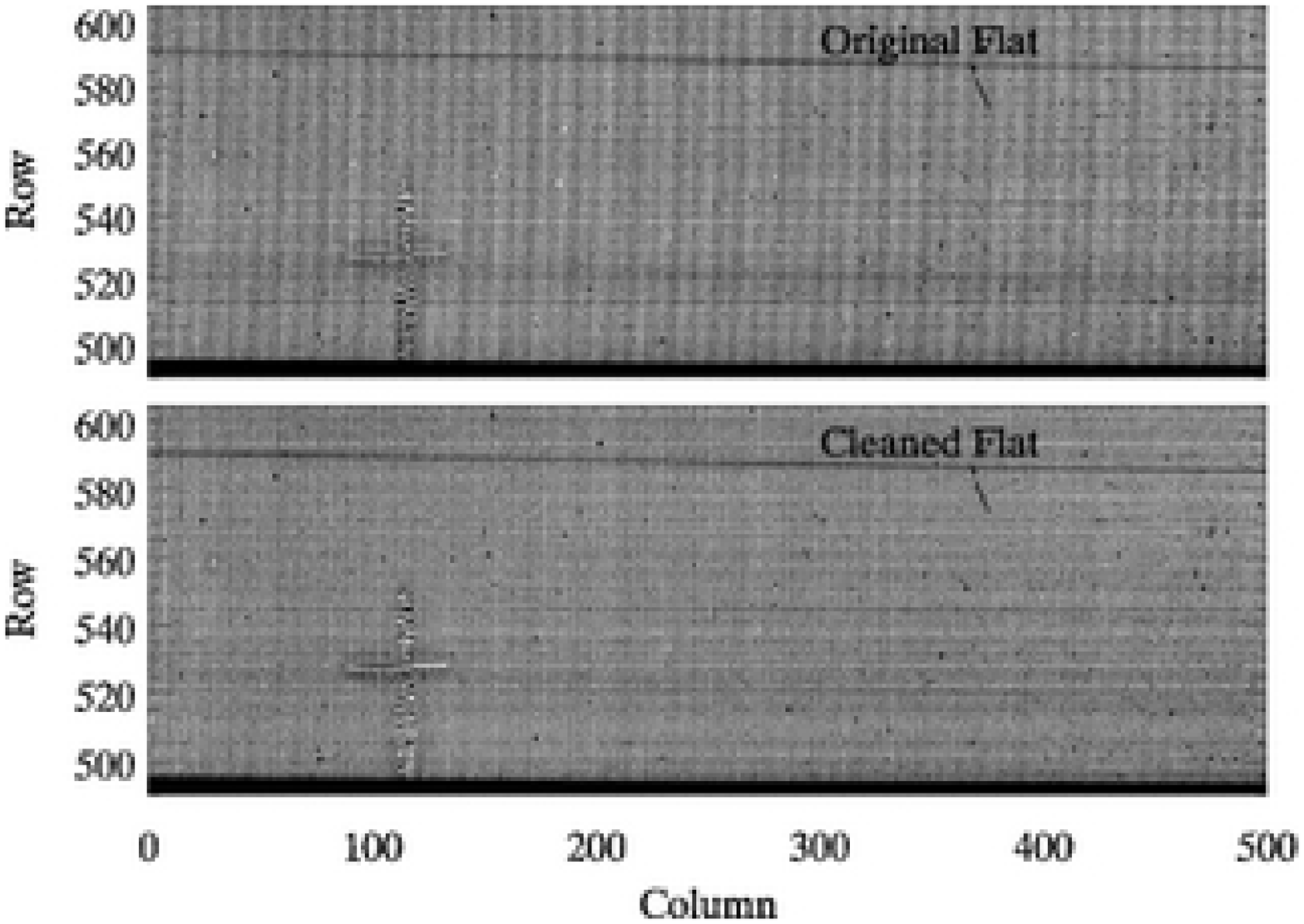}}
\end{ffigure}


\section{\toup{Performance Tests}}\label{noise}

The ratio of fringe pattern amplitude to the pixel-to-pixel variation
(or noise) level varies among different instruments.  We tested the
algorithm's performance at different noise levels by using a synthetic
image consisting of a fringe pattern, a background intensity, and random
noise with a Gaussian distribution that mimics pixel-to-pixel flat-field
variations and photon noise.

The fringe pattern was created using an analytic function that mimics
the oscillating pattern in our example image.  Its functional form is
\begin{equation}
F(x,y) = A  \sin(\nu(x,y) x + \phi(y)),
\end{equation}
where $x$ and $y$ are the position indices in the array, $\phi()$ and
$\nu()$ are linear functions fit to the phase and frequency,
respectively, of our example's fringe pattern, and $A$ is the
amplitude. For these tests we keep the amplitude constant, but there is
no reason for $A$ to be constant in a real image, nor is there any
reason for a non-constant amplitude to adversely affect our algorithm.
The background level is a double-linear function in both $x$ and $y$
directions and has an edge taper.

We define {\it noise strength} as the standard deviation of the Gaussian
noise divided by the standard deviation of the noiseless fringe pattern
($ 2^{-1/2}A$, due to its sinusoidal nature).  Figure \ref{noiseplot}
shows the fraction of remnant fringe after running the algorithm on
simulated data with different fitting functions and varying noise
strength.  The remnant fringe level is not strongly dependent on noise
strength and all methods show very similar behavior with slight
numerical differences when the noise strength is below $\approx 8$.
However, GFC consistently gives the best results in all cases, even
improving when smoothing at high noise levels.  Most of the methods
remove over 95\% of the fringe at noise strength $\approx 0.5$ and over
55\% at noise strength $\approx 9$ (equivalent to Fig. \ref{ex_orig}'s
noise strength).  The lower plot of Fig. \ref{noiseplot} confirms the
intuitive result that the method yields better absolute results for
smaller initial fringe amplitudes.

\begin{ffigure}%
  [Remaining fringe for varying noise strengths.]%
  {Remaining fringe for varying noise strengths.  Only 6 of the 14
    methods are shown above, for clarity.  Omitted methods are similar
    to plotted methods and fall within the range of traces shown.  The
    two nokeep methods give the best results.  In general, all methods
    give similar results for low noise strength, but smoothed nokeep
    methods are better for high noise strength.  Top: Fraction of fringe
    remaining.  Bottom: Absolute fringe remaining when noise level is
    scaled to 1.
    \label{noiseplot}}
  \plotone{\thefigurefile{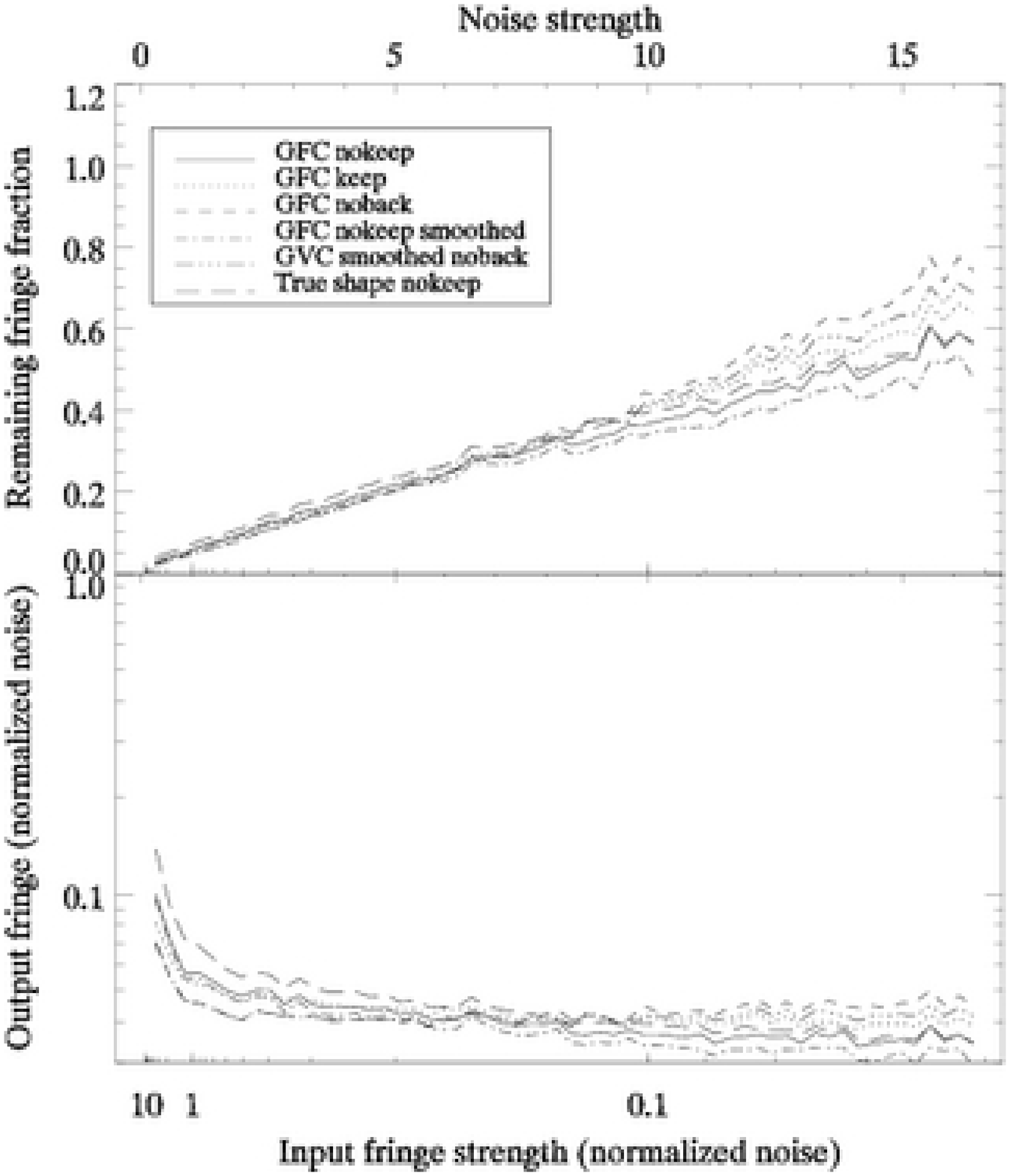}}
\end{ffigure}

Figure \ref{noisebin} shows the effect of varying the bin width.  If the
width is too small when computing the enhanced row, the noise is
insufficiently suppressed.  For low noise, a bin width that is too large
will begin to average out the fringe.

\begin{ffigure}%
  [Remaining fringe for different enhanced--row bin widths]%
  {Remaining fringe for different enhanced-row bin widths.  The panels
    show the same synthetic fringe pattern as Fig.  \ref{noiseplot} at
    three selected noise levels.  Same line styles as in
    Fig. \ref{noiseplot}. \label{noisebin}}
 \plotone{\thefigurefile{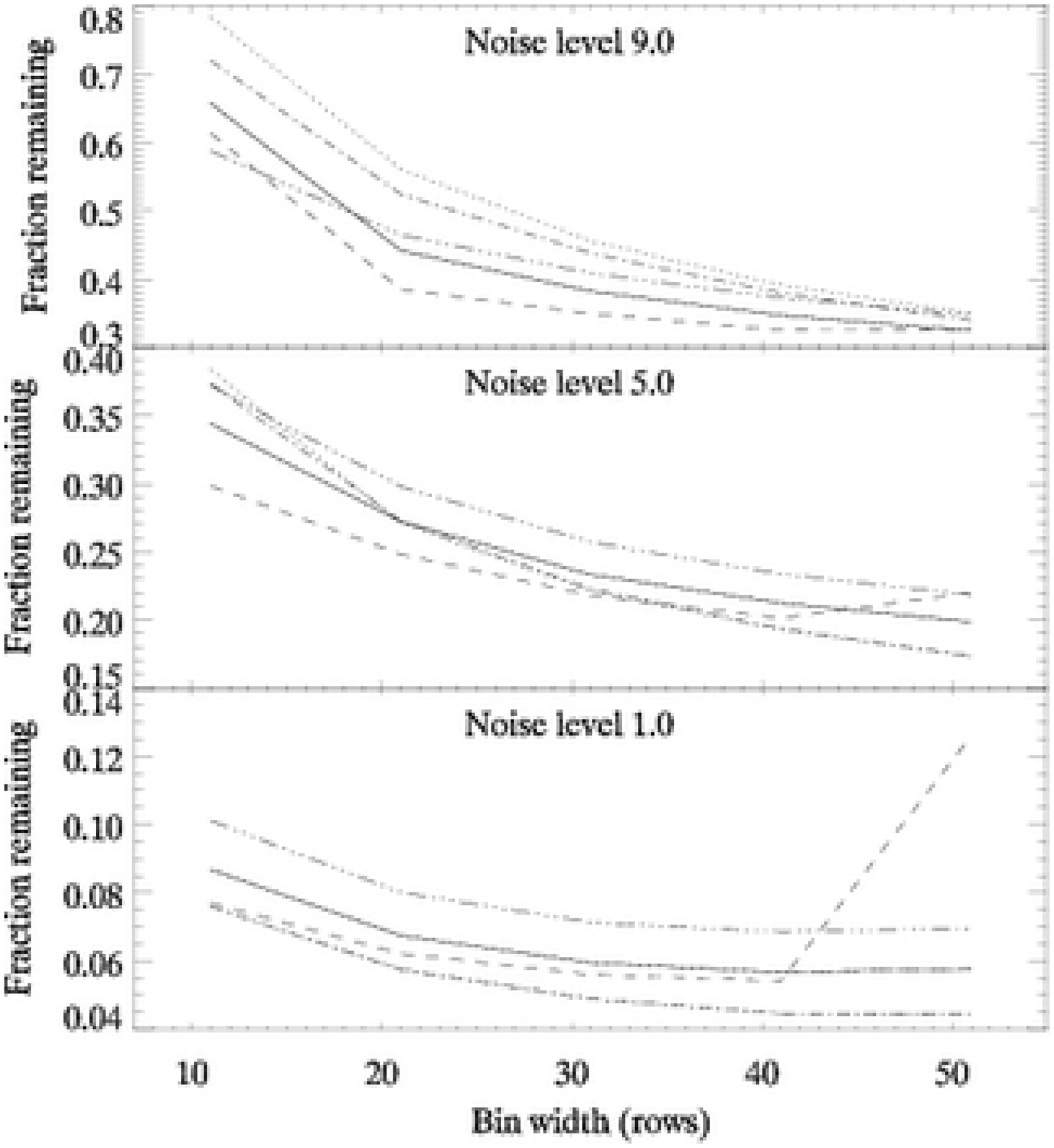}} 
\end{ffigure}

The algorithm is limited by the degree to which the analytic profile
fitting function mimics the data.  Figure \ref{badprof} shows an example
of a difficult profile, which gives very different fits when using the
different fitting functions.  Another source of error is the potential
for the algorithm to miss the correct trace in the presence of high
noise in the wavelet array (Fig. \ref{badtrace}). Also, the
reconstructed fringe pattern is smaller than the input data due to the
factors listed in Step IV. For the example of Fig. \ref{ex_orig}, this
area is $\approx 85$\% of the cropped input image, or over 90\% if only
considering the pixels lost for each surviving row, on average.

\begin{ffigure}%
  [Different fitting methods applied to a profile with a complicated
    shape]%
  {Different fitting methods applied to a profile with a complicated
    shape.  Such shapes are due to noise and are the main limiting
    factor for this algorithm.  This profile comes from the fringe
    transform for row 798, column 627 of our example array.  Crosses
    show the data points, while the solid line is the interpolated
    profile. Other lines are explained in the key.  The profiles are
    only fitted within the local minima at both sides of the reference
    period. \label{badprof}}
  \plotone{\thefigurefile{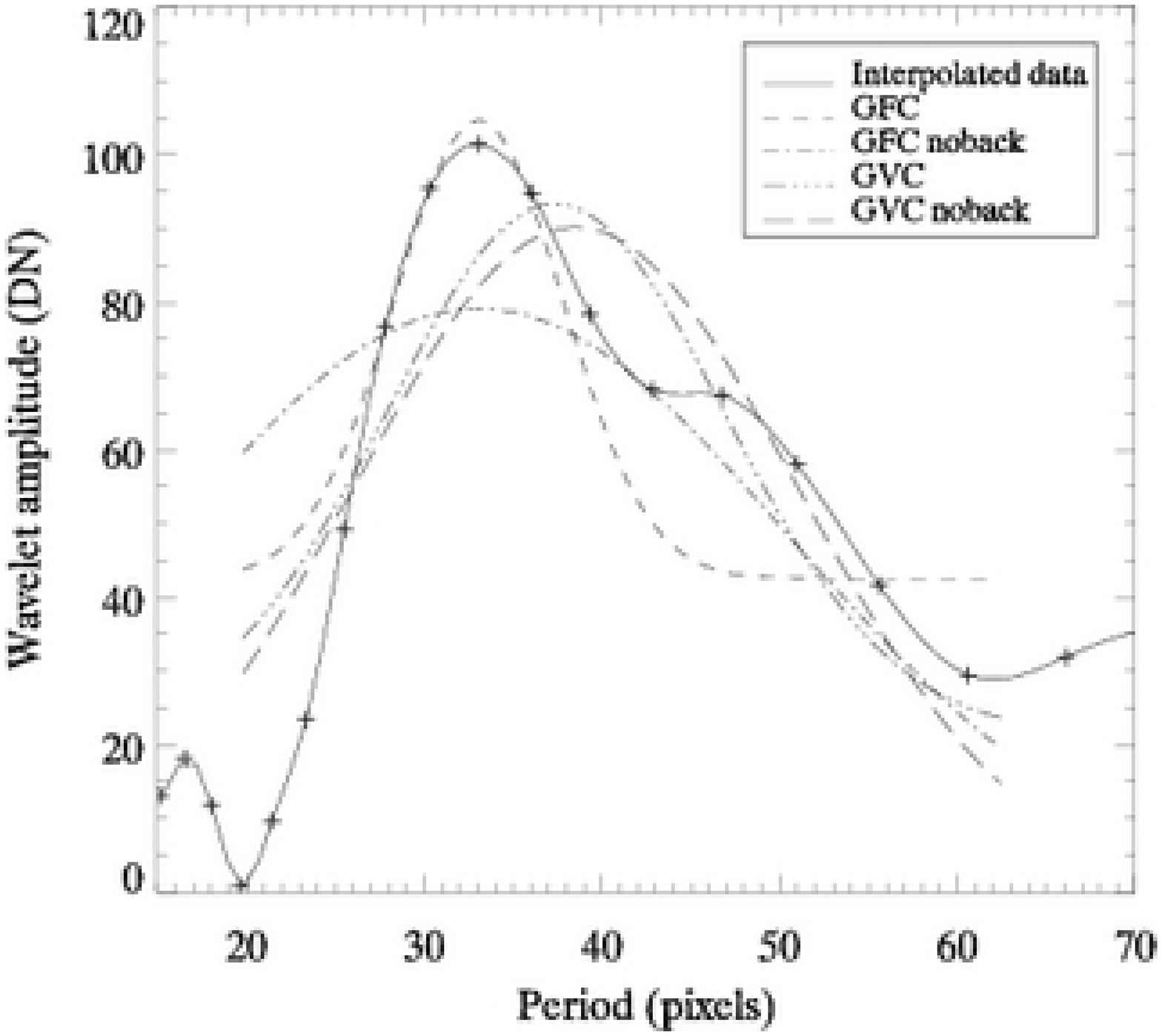}}
\end{ffigure}

\begin{ffigure}%
  [Example of missed trace]%
  {Example of missed trace.  Top: Wavelet array from row 798 of
    Fig. \ref{ex_enhrow}.  Around column 650 the trace goes in the wrong
    direction, towards short period, and disappears around column 720.
    Bottom: Wavelet array from row 799 of Fig. \ref{ex_enhrow}.  Array
    is similar to top plot but now the trace is correct through the last
    column.  Patching in Step III is likely to correct cases like this.
    \label{badtrace}}
  \plotone{\thefigurefile{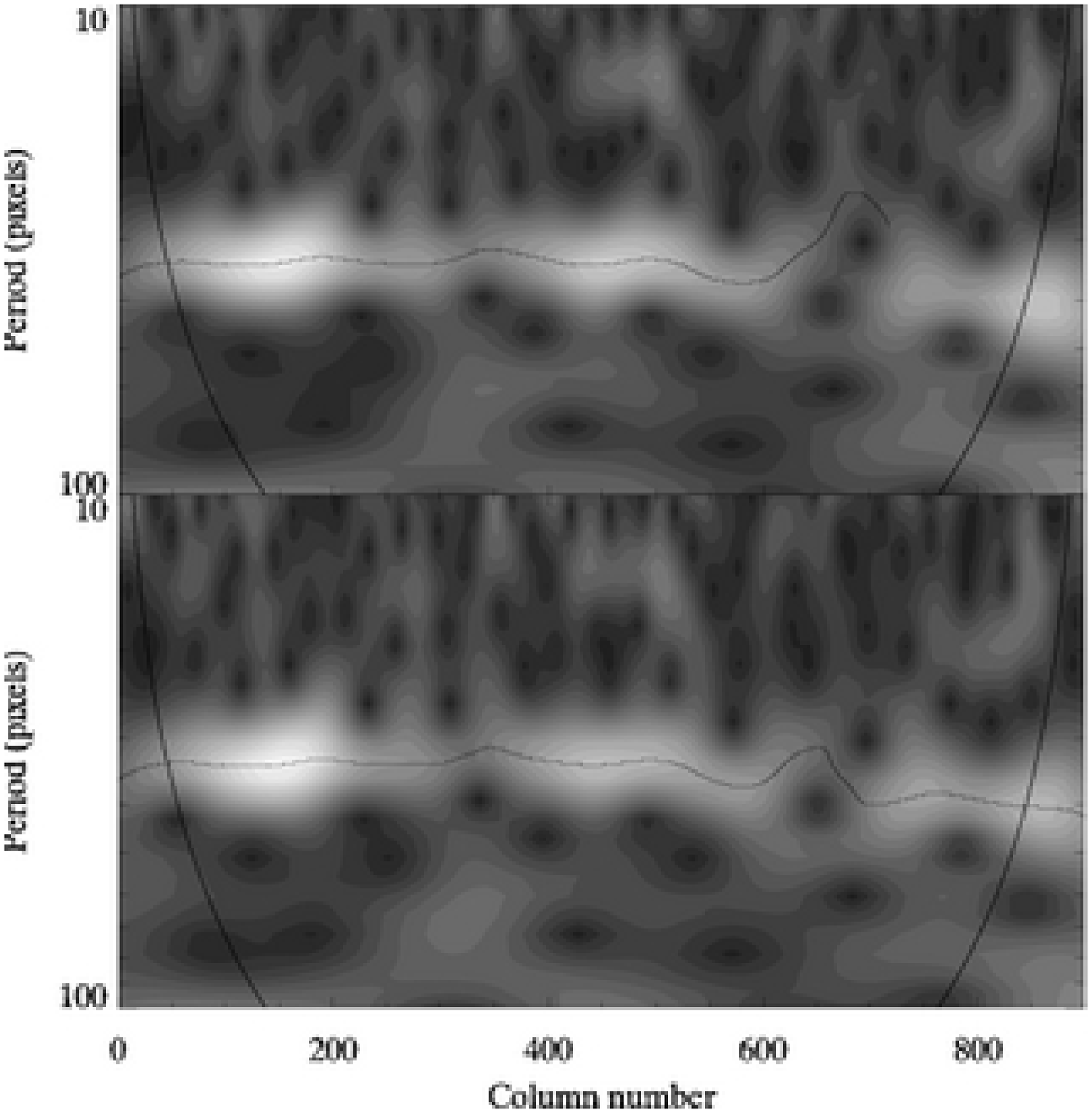}}
\end{ffigure}


\section{\toup{Discussion and Conclusions}}\label{concl}

Seeking a signal from a faint source that is spatially
indistinguishable from a bright source is a long-standing observational
challenge.  Systematic errors that would have been unimportant when
analyzing only the bright source are of concern when considering the
fainter source.  Hence, those errors must be reduced either in the
instrument design or in the data analysis.  To that effect, we have
developed a general algorithm to remove fringe patterns from imaging
data such as flat field frames while preserving other patterns.
Cleaning flat fields is especially useful when the fringe pattern varies
between them and the object data.

Consider the particular example of trying to detect the spectral
modulation of an extrasolar planet as it transits its star using an
instrument like ISAAC at the VLT.  The equivalent noise strength for a
flat field of this instrument is $\approx9$.  On the other hand, no
fringe was detected on the object frames up to a level equivalent to a
noise strength of $\approx 1.5$.  Hence, according to
Fig. \ref{noiseplot}, removing the fringe in the flat fields through our
method would reduce the systematic noise in the data frame by at least
$40\%$.  Considering the flat field intensity, this translates into
residual noise in the data frame $\approx0.25\%$ of the intensity of the
star.  A typical molecular spectral variation is still below that level,
of order $10^{-4}$ times the stellar intensity.  However, it will now be
easier to use the constancy of the planetary signal over many frames to
attain the required sensitivity.

There are three main limitations of this algorithm when applied to a
flat field.  First, the shape of a fringe in wavelet space may be much
more complicated than any reasonable fitting function, resulting in a
partially-corrected fringe.  Second, to be able to follow the trace, the
change in the fringe's period must be smooth.  Finally, there is a
region along the borders where the fringe pattern cannot be recovered.

The algorithm could be improved by finding a parameter-space
interpolation mechanism that would allow defringing of object
frames. Also, a method could be found to fit the entire fringe transform
pattern simultaneously in the 3D wavelet space of row, column, and
period. The 2D wavelet transform may be more appropriate for this
approach.

Our IDL implementation of this algorithm and its documentation appear as
an electronic supplement to this article.  Updated versions are
available on our websites or by email request.


\acknowledgements This material is based upon work supported by the
National Aeronautics and Space Administration under Grant No. NAG5-13154
issued through the Science Mission Directorate.  The example flat field
was obtained from public archives of the European Southern Observatory.
Data presented herein were obtained at the W.M. Keck Observatory from
telescope time allocated to the National Aeronautics and Space
Administration through the agency's scientific partnership with the
California Institute of Technology and the University of California. The
Observatory was made possible by the generous financial support of the
W.M. Keck Foundation.  We also wish to thank the referee for insightful
comments on the manuscript.

\textit{Facilities:} VLT:Antu (ISAAC), Keck:II (NIRSPEC)

%


\bibliographystyle{apj}
\bibliography{/home/pato/inv/common/bibliography}

\end{document}